\documentclass[12pt]{article}
\usepackage{color,amsmath,mathtools,setspace,hyperref,accents,array,tabu,
stackengine,imakeidx,amssymb,graphicx,amsfonts,xcolor,soul}
\usepackage{colortbl}
\usepackage{xcolor}
\usepackage{amsmath,lipsum}
\usepackage{amsfonts}
\usepackage{amssymb,tikz,pgfplots}
\usetikzlibrary{shapes,arrows,snakes}
\usepackage[utf8]{inputenc}
\usepackage[english]{babel}

\DeclarePairedDelimiterX\braket[2]{\langle}{\rangle}{#1 \delimsize\vert #2}
\allowdisplaybreaks
\newcommand\beq{\begin{equation}}
\newcommand\eeq{\end{equation}}
\newcommand\bea{\begin{eqnarray}}
\newcommand\eea{\end{eqnarray}}
\begin{document}
\phantom{xxx} 
\bigskip
\begin{center} 
	{\Large \bf Resolution of SU(3) Outer Multiplicity Problem and the $SU(3)\otimes SU(3)$ Invariant Group $SO(4,2)$} 
\end{center} 
\vskip .5 true cm
\begin{center} 
	{\bf Manu Mathur} \footnote{manu@boson.bose.res.in}, 
	{\bf Atul Rathor} \footnote{atulrathor@bose.res.in}, 
	{\bf T. P. Sreeraj} \footnote{sreeraj.tp@gmail.com, sreerajtp@bose.res.in}
	\vskip 0.6 true cm
	S. N. Bose National Centre for Basic Sciences \\ 
	JD Block, Sector III, Salt Lake City, Calcutta 700098, India.
\end{center} 
\bigskip
\centerline{\bf Abstract}

\noindent We resolve the SU(3) outer multiplicity problem by defining all possible  $SU(3)\otimes SU(3)$ invariant operators in terms of SU(3) Schwinger bosons. We show that the  elementary invariant operators
relevant to the outer multiplicity problem 
 form SO(4,2) algebra. Further, they enable us to construct a family of  operators any one of which can be used to distinguish  repeating representations present in the reduction of the direct product of two SU(3) irreducible representations. \\
\vskip .4 true cm
\section{\bf Introduction}\label{intros} 
A complete labelling of all irreducible representations (IRs) appearing in the reduction of the direct product of two SU(3) IRs ($SU(3)\otimes SU(3) \downarrow SU(3)$) has been a very old and challenging problem \cite{swart,mosh,su3multiplicity}. This problem is usually referred to as the SU(3) 
outer multiplicity or outer degeneracy problem. 
In this paper, we follow a systematic group theoretical approach to address this problem using SU(3) Schwinger bosons. The SU(3) Schwinger bosons, being the most elementary SU(3) operators transforming as the fundamental triplets and anti-triplets, provide a natural framework to resolve this problem (also see \cite{mosh}).  The Schwinger bosons and their simple SU(3) transformation properties enable us to construct all possible  mutually independent operators which are 
invariant under the simultaneous $SU(3) \otimes SU(3)$ transformations. We show that they satisfy SO(4,2) algebra and provide a set of operators which uniquely characterize 
all  IRs present in the reduction of any SU(3) direct product spaces.  Infact, in 1963  
de Swart \cite{swart} had suggested going outside the SU(3) group 
to find such operators. He also suggested the use of symmetry properties of the various irreducible representations to lift the degeneracies (see below).
As the 15 SO(4,2) generators  are the most elementary 
 $SU(3) \otimes SU(3)$ invariant operators, all chiral operators proposed by 
 others \cite{su3multiplicity} in the past as the missing operators are their composites. 
We start with the simplest and widely discussed example of 
SU(3) outer multiplicity problem: 
$$8 \otimes 8 = 1+ 8_a + 8_s + 10 + \overline{10} + 27.$$  
In the above reduction the two octets $8_s$ and $8_a$ are usually differentiated  by their symmetry and antisymmetry  properties respectively under the interchange of the two 
octets on the left hand side. However the above limited characterization under interchange works \cite{su3multiplicity} only when (a) the two direct product IRs are of same dimensions (b)  the multiplicity is 2. In this letter we will use the above example 
to explicitly demonstrate  our technique. We first 
discuss the invariant SO(4,2) operators in terms of the SU(3) Schwinger bosons.  
\section{ $SU(3)$ Schwinger Bosons}

We start with the SU(3) Schwinger boson representation of $SU(3)$ flux operators \cite{georgi}:  
\begin{align}
J^a_{1} &=a^{\dagger}_ {\alpha}\left(\frac{\lambda^a}{2}\right)^{\alpha}_{~ \beta}a^\beta-b_\alpha \left(\frac{\lambda^a}{2}\right)^{\alpha}_{~ \beta}b^{\dagger \beta}, ~~~~
J^a_{2} =c^{\dagger}_ {\alpha}\left(\frac{\lambda^a}{2}\right)^{\alpha}_{~ \beta}c^\beta-d_\alpha \left(\frac{\lambda^a}{2}\right)^{\alpha}_{~ \beta}d^{\dagger \beta}.
\label{su3fo} 
\end{align}
Above $\alpha,\beta =1,2,3$ and $a=1,2,\cdots,8$.  Here, $[a^{\alpha},a^\dagger_\beta] = [b_{\beta},b^{\dagger\alpha}]=[c^{\alpha},c^\dagger_\beta]=[d_{\beta},d^{\dagger\alpha}]=\delta^{\alpha}_\beta $ and $a^\dagger_\alpha,c^\dagger_\alpha$ transforms as triplets and $b^{\dagger\alpha},d^{\dagger\alpha}$ transforms as anti-triplet under the corresponding SU(3) transformations.
We also define the total SU(3) flux operators:
\bea 
J^a \equiv J^a_1+J^a_2. 
\label{tsu3}
\eea 
Each SU(3) IR state is labeled by the eigenvalues of the two Casimir operators and three magnetic operators.
The two Casimir operators for each SU(3) group are given by the number operators $(\hat N_a \equiv a^\dagger \cdot a,\hat N_b\equiv b^\dagger \cdot b)$ and $(\hat N_c\equiv c^\dagger \cdot c,\hat N_d\equiv d^\dagger \cdot d)$.
The eigenvalues of the above  Casimirs or the 
number operators 
will be denoted by $(p_1,q_1)$ and $(p_2,q_2)$ respectively. 
They represent the number of single boxes (triplet) and double boxes (anti-triplet) in  the corresponding Young tableau. 
The three magnetic quantum numbers $i_1,m_1,y_1$ and $ i_2,m_2,y_2$ specify the isospin, magnetic quantum number and hypercharge respectively \cite{swart,mosh}.  
Each of the two SU(3) irreducible representations $|p_1,q_1,i_1,m_1,y_1\rangle$ and $|p_2,q_2,i_2,m_2,y_2\rangle$ is traceless in it's triplet and anti-triplet indices and they satisfy \cite{georgi, su3prep}:
\begin{align} 
(a\cdot b)~ |p_1,q_1,i_1,m_1,y_1\rangle  \equiv 0, ~~~~~~~~
{(c\cdot d)} ~|p_2,q_2,i_2,m_2,y_2\rangle\equiv 0. 
\label{tc2}
\end{align}
  In order to reduce the direct product space into direct sum of IRs, one is required to make a transformation from uncoupled basis to coupled basis. We notice that ten quantum numbers label the uncoupled basis. On the other hand, the coupled states are usually characterized by 9 quantum numbers as $|p_1,q_1,p_2,q_2,p,q,i,m,y\rangle.$ In this labeling scheme, the quantum numbers $(p,q)$ count the numbers of single and double boxes in the coupled Young tableau and are related to the eigenvalues of the quadratic and cubic Casimir operators  for the coupled $SU(3) \otimes SU(3)$ group generated by  (\ref{tsu3}). The quantum numbers $(i,m,y)$ are the eigenvalues of  the total isospin, magnetic and hypercharge operators respectively. Therefore, we need a  $10^{th}$ operator to close the complete set of commuting operators. The eigenvalues of this operator should also differentiate all the repeating IRs present in a direct product.

\section{$SU(3) \otimes SU(3)$ Invariant  SO(4,2) Algebra}

We note that under simultaneous $SU(3)$
transformations generated by $J^a  (= J_1^a+J_2^a)$ in 
(\ref{tsu3}) the 
operators $(a^\dagger_\alpha, c^\dagger_\alpha)$ and 
the operators  $(b^{\dagger\alpha}, d^{\dagger\alpha})$ 
transform as triplets and anti-triplets respectively. 
Therefore, there are all together eighteen\footnote{The remaining eight (cubic) invariants of the  type: $$(\hat \kappa_{\pm}^{(ac;b)},\hat \kappa_{\pm}^{(ac;d)},\hat \kappa_{\pm}^{(bd;a)},\hat \kappa_{\pm}^{(bd;c)}),$$ where $ \hat \kappa_{\pm}^{(ac,b)}=[(a^\dagger \times c^\dagger )\cdot b]~ etc.$, are not considered in (\ref{inv})  because they are not relevant for the  resolution of multiplicity problem as explained in the next Section.
	 Note that these cubic invariants also make the algebra non-linear.}
$SU(3) \otimes SU(3)$ invariant operators 
which are easily constructed as follows:
 \begin{align}
 \Big(\hat{k}^{(ab)}_{\pm},\hat{k}^{(ab)}_{0},\hat{k}^{(cd)}_{\pm},\hat{k}^{(cd)}_{0}, \hat{k}^{(ad)}_{\pm},\hat{k}^{(ad)}_{0},\hat k^{(bc)}_{\pm},\hat k^{(bc)}_{0}\Big),~~\left(\hat \kappa^{(ac)}_{\pm},\hat \kappa^{(ac)}_{0},\hat\kappa^{(bd)}_{\pm},\hat\kappa^{(bd)}_{0}\right)
 \label{inv}
 \end{align}
 where $\left(\hat k^{(ab)}_\pm, \hat k^{(ab)}_0\right)$ and $\left(\hat \kappa^{(ac)}_+,\hat\kappa^{(ac)}_-,\hat\kappa^{(ac)}_0\right)$ are defined as 
 \begin{align}
& \hat k_+^{(ab)} \equiv a^\dagger\cdot b^\dagger, ~~~~~~
 \hat k_-^{(ab)} \equiv a\cdot b,~~~~~~ \hat k_0^{(ab)} \equiv (\hat N_a+\hat N_b +3) \nonumber\\
 & \hat \kappa_+^{(ac)} \equiv a^\dagger \cdot c, ~~~~~~~   \hat \kappa_-^{(ac)} \equiv c^\dagger \cdot a, ~~~~~~~\hat \kappa_0^{(ac)} = (\hat N_a-\hat N_c).
 \label{aabbcc} 
 \end{align}
However, not all the operators in (\ref{aabbcc}) are independent as we can trivially  write the following three identities:
\begin{align} 
\hat\kappa^{(ac)}_0  =  \hat k^{(ab)}_0 - \hat k^{(bc)}_0, \hspace{1.5cm}
\hat\kappa^{(bd)}_0  =  \hat k^{(ab)}_0 - \hat k^{(ad)}_0, \hspace{1.5cm} 
\hat k^{(ab)}_0 + \hat k^{(cd)}_0  =  \hat k^{(ad)}_0+ \hat k^{(bc)}_0.
\end{align}
As a result, we are left with 15 independent $SU(3) \otimes SU(3)$ invariant operators. We now show that  they form SO(4,2) algebra. 
We define  the following two tensor operators: 
\begin{align} 
X^{ \alpha}_{\sigma} = \begin{bmatrix} X^{\alpha}_{\sigma=1} \\ 
	X^{\alpha}_{\sigma=2}\end{bmatrix}\equiv  \begin{bmatrix} a^\alpha  \\ 
	c^\alpha\end{bmatrix}, \hspace{3cm}
Y^{\sigma}_{\alpha} = \begin{bmatrix} Y_{\alpha}^{\sigma=1} \\ 
	Y_{\alpha}^{\sigma=2}\end{bmatrix} \equiv  \begin{bmatrix} b_\alpha  \\ 
	d_\alpha\end{bmatrix}.
\label{so42nw} 
\end{align} 
In (\ref{so42nw}), $\sigma =1,2; ~\alpha =1,2,3$. We  now construct the 15 $SU(3) \otimes SU(3)$ invariant SO(4,2) generators $L_{\mu\nu} = -L_{\nu\mu}, 
\mu,\nu =1,2,\cdots ,6$  as follows \cite{wyebourn}: 
 \begin{align}
 \begin{split}\label{so42a}
&\hat L_{ij} \equiv \frac{1}{2}\epsilon_{ijk}~ Tr\left(X^\dagger \sigma^k X  + Y^\dagger \sigma^k Y\right) \hspace{1.5cm} ~ \hat L_{i4} \equiv 
-\frac{1}{2} Tr\left(X^\dagger \sigma^i X  - Y^\dagger \sigma^i Y\right) \\
&\hat L_{i5} \equiv -\frac{1}{2} Tr\left(X^\dagger \sigma^i \tilde Y^\dagger  - \tilde Y \sigma^i X\right)\hspace{1.7cm}~~
\hat L_{i6} \equiv -\frac{i}{2} Tr\left(X^\dagger \sigma^i \tilde  Y^\dagger  + \tilde Y \sigma^i X\right) \\
& \hat L_{46}\equiv  \frac{1}{2}Tr
\left(X^\dagger \cdot \tilde Y^\dagger + X \cdot \tilde Y\right) \hspace{2.1cm}  ~~ \hat L_{45}
\equiv  -\frac{i}{2} Tr
\left(X^\dagger \cdot \tilde Y^\dagger - X \cdot \tilde Y\right) \\ & \hspace{4cm} \hat L_{56} \equiv \frac{1}{2} Tr \left( X^\dagger \cdot X + Y^\dagger \cdot Y+2 \right).
\end{split} 
\end{align} 
In (\ref{so42a}]) $i,j,k =1,2,3$ and  
the traces are over the SU(3) indices $\alpha,\beta=1,2,3$. 
 make all 15 $\hat L_{\mu\nu}$  invariant under any simultaneous $SU(3)\otimes SU(3)$ transformation.
The SO(4,2) algebra can be easily verified: 
\begin{align} 
\left[\hat L_{\mu\nu}, \hat L_{\rho\sigma}\right] = i \left(g_{\mu\rho} \hat L_{\nu\sigma}+ g_{\nu\sigma} \hat  L_{\mu\rho} + g_{\mu\sigma} \hat L_{\rho\nu}+ g_{\nu\rho} \hat L_{\sigma\mu}\right).
\label{so42al} 
\end{align}
In (\ref{so42al}) $g_{\mu\nu}$ represents the diagonal metric $(+~+~+~+~-~-)$. Note that the $L_{\mu\nu}$ operators are linear combinations of the operators in (\ref{inv}). Further, all operators appearing in (\ref{inv}) and (\ref{aabbcc}) can be constructed in terms of SO(4,2) generators $L_{\mu\nu}$. Their  $SU(3)\otimes SU(3)$ invariance can also be easily checked:  
 \bea 
\hspace{4cm} \left[\hat J^a, \hat L_{\mu\nu}\right]=0,\hspace{1cm} a=1,2,\cdots ,8; ~~\mu,\nu =1,2,\cdots ,6. 
\label{su342} 
\eea 
 The three SO(4,2) Casimirs are: 
\begin{align}
\hat {\cal C}_2 =\hat L_{\mu\nu}\hat L^{\mu\nu} \hspace{1.5cm} \hat {\cal C}_3=\epsilon_{\mu\nu\rho\sigma\delta\kappa}\hat L^{\mu\nu}\hat L^{\rho\sigma}\hat L^{\delta\kappa} \hspace{1.5cm} \hat {\cal C}_4=\hat  L_{\mu\nu}\hat L^{\nu\rho}\hat L_{\rho\sigma}\hat L^{\sigma\mu}.
\end{align}
Where $\hat L^{\mu\nu} =g^{\mu \rho}g^{\sigma\nu}\hat L_{\rho\sigma}$. Note that $\hat {\cal C}_2, \hat {\cal C}_3$ and $\hat {\cal C}_4$ commute with all invariants  and hence also with the invariant constraints (\ref{tc2}): 
\begin{equation}
\left[\hat {\cal C}_\alpha, \hat L_{\mu\nu}\right]=0,\qquad \left[\hat{\cal C}_\alpha, \hat k_{-}^{(ab)}\right]=0,\qquad \left[\hat {\cal C}_\alpha,\hat k_{-}^{(cd)}\right]=0.
\end{equation}

\section{Resolution of the problem}\label{4}

We require a complete set of commuting operators (CSCO) 
containing 10 hermitian operators whose eigenvalues parametrize all coupled states uniquely. Three $SU(3) \otimes SU(3)$ magnetic operators $\hat I^2, \hat I_3,\hat Y$ constructed out of $J^a$ in (\ref{tsu3}) and the four Casimirs of the two SU(3) $ \hat N_a,\hat N_b,\hat N_c,\hat N_d$ provide seven of them. Therefore, we require three invariant operators constructed out of $\hat L_{\mu\nu}$ which commutes with each other and also with the 4 number operators. The three Casimirs $\hat {\cal C}_2,\hat {\cal C}_3$ and $\hat {\cal C}_4$ of SO(4,2) are the most natural choices as they also commute with the  invariant operators $(\hat k_-^{(ab)}, \hat k_-^{(cd)})$ and thus preserving the constraints $\hat k_-^{(ab)}\approx 0$ and  $\hat k_-^{(cd)}\approx 0$
 in (\ref{tc2}). However, within this  constrained Hilbert space, $\hat {\cal C}_4$ is not independent
 of $\hat {\cal C}_2$ and $\hat {\cal C}_3$. Therefore, we define the last missing operator in the CSCO in the following 2 steps: 
 \begin{enumerate}  
  \item 
The most general operator\footnote{Any invariant operator constructed out of the cubic invariants of the type $\hat \kappa_{\pm}^{(ac,b)}$  which commutes with $\hat {\cal C}_2, \hat {\cal C}_3$ and $\hat N_a,\hat N_b,\hat N_c,\hat N_d$ can written in terms of $SO(4,2)$ invariants using the identity  
$  \epsilon_{ijk}\;\epsilon_{klm}  =\;\delta _{il}(\delta _{jm}\delta _{kn}-\delta _{jn}\delta _{km})-\;\delta _{im}(\delta _{jl}\delta _{kn}-\delta _{jn}\delta _{kl})+\delta _{in}(\delta _{jl}\delta _{km}-\delta _{jm}\delta _{kl})$. This is the reason why we could ignore the cubic invariants to get the $SO(4,2)$ algebra in the last section.} 
constructed out of Schwinger bosons which commutes with the nine operators in the set $\left(\hat N_a,\hat N_b,\hat N_c,\hat N_d, \hat {\cal C}_2, \hat {\cal C}_3, \hat I^2, \hat I_3,\hat Y \right)$
is given by: 
\begin{align}
\hat {\cal C}_4'=\lambda_1(a^\dagger \cdot c)(c^\dagger \cdot a) 
+\lambda_2 (b^\dagger \cdot d)(d^\dagger \cdot b)
+\lambda_3 (a^\dagger \cdot d^\dagger)(a \cdot d)+\lambda_4(b^\dagger \cdot c^\dagger)(b \cdot c).
\end{align}
\item  In order to preserve the constraints $\hat k_-^{(ab)}\approx 0$ and  $\hat k_-^{(cd)}\approx 0$ to 
retain the tracelessness properties of the two SU(3) IRs, 
we replace all SU(3) Schwinger bosons by the corresponding SU(3) irreducible Schwinger bosons \cite{su3prep} to get: 
\begin{align}
\hat {\cal C}_4'=\lambda_1(A^\dagger \cdot C)(C^\dagger \cdot A) 
+\lambda_2 (B^\dagger \cdot D)(D^\dagger \cdot B)
+\lambda_3 (A^\dagger \cdot D^\dagger)(A \cdot D)+\lambda_4(B^\dagger \cdot C^\dagger)(B \cdot C). 
\label{su3isb} 
\end{align}
\end{enumerate} 
In (\ref{su3isb}) $A^{\dagger}_{\alpha}, B^{\dagger\alpha}, C^{\dagger}_{\alpha}, D^{\dagger \alpha}$ are the SU(3) irreducible Schwinger bosons  defined as \cite{su3prep}: 
\begin{align}
\begin{split}\label{defirrsh}
A^{\dagger}_{\alpha}=&a^{\dagger}_{\alpha}-\frac{1}{\hat{N}_{a}+\hat N_{b}+1}\hat{k}_+^{(ab)}b_\alpha\qquad \qquad C^{\dagger}_{\alpha}=c^{\dagger}_{\alpha}-\frac{1}{\hat{N}_{c}+\hat N_{d}+1}\hat{k}_+^{(cd)}d_\alpha\\
B^{\dagger\alpha}=&b^{\dagger\alpha}-\frac{1}{\hat{N}_{a}+ \hat N_{b}+1}\hat{k}_+^{(ab)}a^\alpha\qquad \qquad D^{\dagger\alpha}=d^{\dagger\alpha}-\frac{1}{\hat{N}_{c}+ \hat N_{d}+1}\hat{k}_+^{(cd)}c^\alpha
\end{split}
\end{align}
We choose the simplest form for $\hat {\cal C}_4'$ by taking $\lambda_1=1$ and $\lambda_2=\lambda_3=\lambda_4=\lambda_5=\lambda_6=0$.
The CSCO $\left(\hat N_a,\hat N_b,\hat N_c,\hat N_d, \hat {\cal C}_2, \hat {\cal C}_3, \hat {\cal C}^\prime_4,\hat I^2, \hat I_3,\hat Y\right)$ can be diagonalized to characterize the coupled basis vectors uniquely.
We now illustrate our procedure using the $8 \otimes 8$ example  discussed in the introduction.  
The $|8\rangle$ and $|8'\rangle$ states\footnote{$|8\rangle_s$, $|8\rangle_a$ mentioned in the introduction are the two octets $(A^\dagger \cdot D^\dagger)C^{\dagger }_{\alpha} B^{\dagger\beta}|0\rangle \pm (C^\dagger\cdot B^\dagger)  A^{\dagger}_{ \alpha} D^{\dagger \beta}|0\rangle$. They are  symmetric and anti-symmetric under  the exchange $a^\dagger \leftrightarrow c^\dagger,~  b^\dagger \leftrightarrow d^\dagger$ respectively.} in the direct sum can be written in terms of irreducible Schwinger bosons as 
{\footnotesize \begin{align}
|8\rangle ^\beta_\alpha~\equiv ~& (A^\dagger \cdot D^\dagger)  C^{\dagger }_{\alpha} B^{\dagger \beta}|0\rangle-4(B^\dagger\cdot C^\dagger)  A^{\dagger}_{ \alpha} D^{\dagger \beta}|0\rangle\nonumber\\
|8'\rangle^\beta_\alpha ~ \equiv ~ &  (A^\dagger\cdot D^\dagger)  C^{\dagger}_{ \alpha} B^{\dagger \beta}|0\rangle-\tfrac{1}{2}(B^\dagger\cdot C^\dagger)  A^{\dagger}_{\alpha} D^{\dagger \beta}|0\rangle.
\label{x14} 
\end{align}}
The action of $\hat {\cal C}_4'$ is given by:
\begin{align}
\hat {\cal C}_4' ~|8\rangle^\beta_\alpha=  \bigg(\frac{3}{4}\bigg)|8\rangle^\beta_\alpha ,\hspace{1cm}
\hat {\cal C}_4' ~|8'\rangle^\beta_\alpha=  (0)|8'\rangle^\beta_\alpha. \nonumber\\
\end{align}
Thus the two octet states defined in (\ref{x14}) have different eigenvalues with respect to the new Casimir operator $\hat {\cal C}_4^\prime$. 

\section{Conclusions} 

In this work  we have constructed the minimal and complete set of algebraically independent $SU(3) \otimes SU(3)$ invariant operators satisfying SO(4,2) algebra.  These invariant operators, in turn, help us define the complete set of commuting operators in the SU(3) direct product space solving the outer multiplicity problem. 
The present Schwinger boson approach can be directly generalized to resolve outer multiplicity problem for all SU(N)  by simply working with SU(N) Schwinger bosons and constructing the corresponding $SU(N) \otimes SU(N)$ invariant group. These SU(N) results will be published elsewhere.  The computation of all SU(3) Clebsch Gordan coefficients in the present basis will also be discussed in the next work.
	
\end{document}